\documentstyle{article}

\def\ba{\begin{array}}
	 \def\ea{\end{array}}
	 \def\be{\begin{equation}}
	 \def\ee{\end{equation}}

	 \def\D{{\Delta }}
	 \def\P{{\Phi}}
	 \def\p{{\phi}}
	 \def\si{{\psi}}
	 \def\d{{\partial}}
	 \def\R{{\bf R}}
	 \def\be{\begin{equation}}
	 \def\ee{\end{equation}}
	 \def\bea{\begin{eqnarray}}
	 \def\eea{\end{eqnarray}}
	 \def\ba{\begin{array}}
	 \def\ea{\end{array}}
	 \def\s{\sum}

	 \def\Si{\Psi}

	 \def\D{\Delta}
	 \def\b{\beta}

	 \def\cs{\cdots}
	 \def\e{\epsilon}
	 \def\g{\gamma}

\def\W{{\cal W}}

\def\de#1{\Delta_{#1}}

\def\I{\rm {I\kern-.3em I}}
\def\C{\rm {I\kern-.520em C}}
\def\R{\rm {I\kern-.3em R}}
\def\CZ{\rm {Z\kern-.4em Z}}

\def\unit{\rm {1\kern-.4em 1}}

\def\kk{{\bf k}}
\def\dis{\displaystyle}
\begin{document}
\begin{titlepage}
\vspace{-10mm}
\begin{flushright}
	     Mar. 1997\\
					     \end{flushright}
\vspace{12pt}
%%%%%%%%%%%%%%%%%%%%  The title page   %%%%%%%%%%%%%%%%%%%%%%%%%%%%%%%%%%%%%
%\hfill{}
%\vbox{
%    \halign{#\hfil         \cr
%            hep-th/9610168 \cr\noalign{\vskip -0.5cm}
%            IPM 96-        \cr\noalign{\vskip -0.5cm}
%            TUDP 96      \cr\noalign{\vskip -0.5cm}
%            IASBS 96     \cr\noalign{\vskip -0.5cm}
%           } % end of \halign
%      }  % end of \vbox
\vskip .5 mm
\leftline{ \Large \bf
	  The Logarithmic Conformal Field Theories}
\leftline{ \bf
	  M. R. Rahimi Tabar ${}^{1,2,*}$,
	  A. Aghamohammadi ${}^{1,3}$, and
	  M. Khorrami ${}^{1,4,5}$
	  }

\leftline{ \bf
	  }
\vskip .02 mm
{\it
  \leftline{ $^1$ Institute for Studies in Theoretical Physics and
	    Mathematics, P.O.Box  5531, Tehran 19395, Iran. }
 \leftline{ $^2$ Department of Physics, University of Science and 
	    Technology, Narmak, Tehran 16844, Iran}
  \leftline{ $^3$ Department of Physics, Alzahra University,
	     Tehran 19834, Iran. }
  \leftline{ $^4$ Department of Physics, Tehran University,
	     North Kargar Ave. Tehran, Iran. }
  \leftline{ $^5$ Institute for Advanced Studies in Basic Sciences,
	     P.O.Box 159, Gava Zang, Zanjan 45195, Iran. }
  \leftline{ $^*$ Rahimi@netware2.ipm.ac.ir}
  }
\begin{abstract}
We study the correlation functions of logarithmic conformal field theories. 
First, assuming conformal invariance, we explicitly calculate two-- and 
three-- point functions. This calculation is done for the general case of 
more than one logarithmic field in a block, and more than one set of 
logarithmic fields. Then we show that one can regard the logarithmic 
field as a formal derivative of the ordinary field with respect to its 
conformal weight. This enables one to calculate any $n$-- point function 
containing the logarithmic field in terms of ordinary $n$--point functions.
At last, we calculate the operator product expansion (OPE) coefficients of 
a logarithmic conformal 
field theory, and show that these can be obtained from the corresponding 
coefficients of ordinary conformal theory by a simple derivation. \\
PACS number 11.25.Hf \\
Keyword, Conformal Field Theory
\end{abstract}
\vskip 10 mm
\end{titlepage}
%%%%%%%%%%%%%%%%%%%%  The body of the paper            %%%%%%%%%%%%%%%%%%%%%
{\section {Introduction}}
It has been shown by Gurarie \cite{gu}, that conformal field theories
(CFTs) whose correlation functions exhibit logarithmic behaviour, can be 
consistently defined and in the OPE of two given local
fields which has at least two fields with the same conformal dimension, one
may find some operators with a special property, known as logarithmic
operators. As discussed in \cite{gu}, these operators with the ordinary
operators form the basis of the Jordan cell for the operators $L_i$.
In some interesting physical theories, for example the WZNW model on the 
$GL(1,1)$ super-group \cite{srz}, and edge excitation in fractional quantum 
Hall effect \cite{5}, one can  naturally find logarithmic terms in 
correlators. Recently the role of logarithmic operators have been 
considered in study of some physical problems such as 2D-magnetohydrodynamic 
turbulence \cite{rr1,rr2,rr3},  2D-turbulence \cite{flo,rahim1},
$c_{p,1}$ models \cite{cp1,fl}, gravitationally dressed CFT's \cite{bk}, 
and some critical disordered models \cite{10,d14}. They play  a role in the 
so called unifing $\W$ algebra \cite{W} and in the description of 
normalizable zero modes for string backgrounds \cite{b16,b17}.

The basic properties of logarithmic operators are that,
they form a part of the basis of the Jordan cell
for $L_i$'s and in the correlator of such fields there
is a logarithmic singularity \cite{gu,10}.
It has been shown that in rational minimal models such a situation, 
i.e. two fields with the same dimensions, doesn't occur \cite{rr2}.
The modular invariant partition functions for 
$c_{eff}=1$ and the fusion rules of logarithmic conformal field 
theories (LCFT) are considered in \cite{fl,fufl}.

In this paper, we  study the correlation functions of logarithmic 
conformal field theories (LCFT's). Assuming conformal invariance, we obtain 
all two- and three-point functions. 
This calculations have been already done for the case where the Jordanian
cell is two dimensional \cite{gu,10}.
The  key observation in this point 
is that, one can regard logarithmic fields {\it formally} as the derivative 
of ordinary fields with respect to their conformal weight
and use this effectively to obtain logarithmic three- and more-point
functions from ordinary ones. We think 
that many other results, if not all of them, for LCFT's
can also be obtained   by this technique from ordinary 
conformal field theories. 
We show that any $n$-point 
function for these theories can be obtained
from their analouges in the ordinary conformal field theories. 
These results are then extended to the case of more than two fields in a 
Jordan cell, and more than one Jordan cell. At last we give the OPE 
coefficients of a LCFT with a two-dimensional Jordan cell, in terms of the 
OPE coefficients of the corresponding CFT, and then generalize it to the 
case of a more dimensional Jordan cell.
\vspace{0.5cm}

\section{The Correlation Functions of a LCFT}
In an ordinary conformal field theory, primary fields are the highest weights 
of the representations of the Virasoro algebra. The operator product 
expansion that defines a primary field  $\P (w,\bar w)$ is \cite{bpz}
\be \label{*}
T(z) \P_i (w,\bar w)={\D_i \over (z-w)^2}\P_i (w, \bar w)+{1\over (z-w)}
\d _w \P_i (w,\bar w)
\ee
\be
T(\bar z) \P_i (w,\bar w)={\bar \D_i \over (\bar z-\bar w)}\P_i
(w, \bar w)+{1\over (\bar z-\bar w)}
\d _{\bar w} \P_i (w,\bar w)
\ee
where $T(z):= T_{zz}(z)$ and $\bar T(\bar z):= T_{\bar z \bar z}(\bar z)$.  
The primary fields are those which transform under $z\to f(z)$ and
$\bar z\to \bar f(\bar z)$ as: 
\be \label{1} 
\P_i(z,\bar z)\to {\P}'_i(z,\bar z)=({\d f^{-1}\over \d z})^{\D_i}
({\d {\bar f}^{-1}\over \d \bar z})^{\bar \D_i}
\P_i(f^{-1}(z),\bar f^{-1}(\bar z))
\ee
One can write equation ( \ref{*}) in terms of the  components of 
Laurent expansion of $T(z)$ , $L_n$'s,
\be \label{2}
[L_n,\P_i(z)]=z^{n+1}\d_z\P_i+(n+1)z^n\D_i \P_i
\ee
One can regard $\D_i$'s as the diagonal elements of a diagonal matrix $D$, 
\be 
[L_n,\P_i(z)]=z^{n+1}\d_z\P_i+(n+1)z^nD_i^j \P_j
\ee
One can however, extend the above relation for any matrix $D$, which is not 
necessarily diagonal. This new representation of $L_n$  also satisfies the 
Virasoro  algebra for any arbitrary matrix $D$ . Because we have not 
altered the first term in the right hand side of the equation (\ref{1}), 
this is still a conformal transformation.
By a suitable change of basis, one can make  $D$  diagonal or Jordanian. 
If it becomes diagonal, the field theory is nothing 
but the ordinary conformal field theory. The general case is that there 
are some Jordanian blocks in the matrix $D$. The latter is the 
case of a LCFT. Here, there arise some other fields which do not transform
like ordinary primary fields, and are called quasi-primary fields \cite{gu}.
For the simplest case, consider a two-dimensional Jordan cell.
The fields $\P$ and $\Si$ satisfy 
\be \label{10} 
[L_n,\P(z)]=z^{n+1}\d_z\P+(n+1)z^n\D \P
\ee
and
\be \label{11}
[L_n,\Si(z)]=z^{n+1}\d_z\Si+(n+1)z^n\D \Si+(n+1)z^n \P,
\ee
and they transform as below
\be \label{12} 
\P(z)\to ({\d f^{-1}\over \d z})^{\D}\P(f^{-1}(z))
\ee
\be \label{13} 
\Si(z)\to ({\d f^{-1}\over \d z})^{\D}[ \Si(f^{-1}(z))+
\log ({\d f^{-1}(z)\over \d z}) \P(f^{-1}(z)]
\ee
Note that we have considered only the chiral fields. The logarithmic fields,
however cannot be factorized to the left- and right-handed fields.
For simplicity we derive  the results for chiral fields. The corresponding
results for full fields are simply obtained by changing
\be
z^{\D}\to z^{\D}{\bar z}^{\bar \D}
\ee
and 
\be 
\log z\to \log \vert z\vert^2
\ee
Now compare  the relations (\ref{10}, \ref{12}) and (\ref{11}, \ref{13}); 
one can assume the field $\Si$ as  the derivation of the field $\P$ with 
respect to its conformal weight, $\D$. This fact will be  
effectively used throughout this paper.

Now let us consider the action of M\"obius generators $(L_0,L_{\pm})$ on  
the correlation functions. Whenever the field $\Si$ is absent, the 
{\it form} of the 
correlators is the same as ordinary conformal field theory. By the term 
{\it form} we mean that some of the constants which cannot be determined 
in the ordinary conformal field theory may be fixed in the latter case.
Now we want to compute correlators containing the field $\Si$. At first 
we should compute the two-point functions. The two-point  functions
of the field $\P$ is as below
\be 
<\P(z) \P(w)>={c\over (z-w)^{2\D}}
\ee
In the ordinary conformal field theory the constant $c$ cannot be determined  
only with assuming conformal invariance; to obtain it, one should know for 
example the stress-energy tensor, although for $c\neq 0$ one can set it equal 
to one by renormalizing the field. Assuming the conformal invariance of the
two-point function $<\Si(z) \P(w)>$, means that acting the set 
$\{ L_0,L_{\pm 1}\}$ on the correlator yeilds zero. Action of $L_{-1}$ 
ensures that the correlator depends only on the $z-w$. the relations for 
$L_{+1}$ and $L_0$ are as below
\be
[z^2\d_z+w^2\d_w +2\D(z+w)]<\Si(z)\P(w)>+2z <\P(z)\P(w)>=0
\ee
\be
[z\d_z+w\d_w +2\D]<\Si(z)\P(w)>+ <\P(z)\P(w)>=0 
\ee
Consistency of these two equations for any $z$ and $w$, fixes $c$ to be zero.
Then, solving the above equation for $<\Si(z) \P(w)>$ leads to
\be 
<\P(z) \P(w)>=0,\qquad <\P(z) \Si(w)>={a\over (z-w)^{2\D}}
\ee
Now assuming the conformal invariance of the two-point function 
$<\Si(z) \Si(w)>$, gives us a set of partial differential equation. Solving 
them, we obtain
\be 
<\Si(z) \Si(w)>={1\over (z-w)^{2\D}}[b-2a \log (z-w)]
\ee
These correlations have been obtained in \cite{gu,10}, by assuming the 
consistency of some four point functions. In fact, in \cite{gu} the two 
point functions $\Si\P$ and $\Si\Si$ were obtained using the four point 
function and the assumption that there exists a term with a logarithmic 
factor in the OPE of certain fields. In \cite{10}, using the same 
assumption, it is shown that the correlator $\P\P$ is zero, and some three 
point functions are calculated.

Now we extend the above results to the case where Jordanian block is 
$n+1$-dimensional. So there is $n+1$ fields with the same weight $\D$.
\be 
[L_n,\P_i(z)]=z^{n+1}\d_z\P_i+(n+1)z^n\D \P_i+(n+1)z^n \P_{i-1},
\ee
where $\P_{-1}=0$. 
All we use is the conformal invariance of the theory. From the above fields, 
only  $\P_0$ is  primary.
Acting $L_{-1}$ on any two-point function of these fields, shows that
\be 
<\P_i(z) \P_j(w)>=f_{ij}(z-w). 
\ee
Acting $L_0$ and $L_{+1}$, leads to
\be 
<[L_0,\P_i(z) \P_j(0)]>=(z\d_z +2\D)<\P_i(z) \P_j(0)>+<\P_{i-1}(z) \P_j(0)>
+<\P_i(z) \P_{j-1}(0)>=0
\ee
\be 
<[L_{+1},\P_i(z) \P_j(0)]>=(z^2\d_z +2z\D)<\P_i(z) \P_j(0)>+
2z<\P_{i-1}(z) \P_j(0)> =0.
\ee
Then it is easy to see that 
\be \label{14}
<\P_{i-1}(z) \P_j(0)>=<\P_i(z) \P_{j-1}(0)>.
\ee
Using $\P_{-1}=0$ and the above equation, gives us the following 
two-point functions.
\be 
<\P_i(z) \P_j(w)>=0 \qquad {\rm for }\qquad i+j<n
\ee
Now solving the Ward identities for $<\P_0(z) \P_n(w)> $ among 
with the relation (\ref{14}), leads 
to
\be
<\P_i(z) \P_{n-i}(w)>=<\P_0(z) \P_n(w)>=a_0 (z-w)^{-2\D}.
\ee
The form of the correlation function $<\P_1(z) \P_n(w)>$ is as below
\be
<\P_1(z) \P_n(w)>=(z-w)^{-2\D}[a_1+b_1\log (z-w)],
\ee
but the conformal invariance fixes $b_1$ to be equal to $-2a_0$. So
\be
<\P_i(z) \P_{n+1-i}(w)>=<\P_1(z) \P_n(w)>=(z-w)^{-2\D}[a_1-2a_0\log (z-w)]
\quad {\rm for }\quad i>0
\ee
Repeating this procedure for the two-point functions of the other fields 
$\P_i$ with $\P_n$, and knowing that they are in the following form
\be
<\P_i(z) \P_n(w)>=(z-w)^{-2\D}\sum_{j=0}^i a_{ij}(\log(z-w))^j,
\ee
gives 
\be
\sum_{j=1}^i ja_{ij}(\log(z-w))^{j-1}+2\sum_{j=0}^{i-1} a_{i-1,j}
(\log(z-w))^j =0
\ee
or
$$(j+1)a_{i,j+1}+2a_{i-1,j}=0$$
So
\be
a_{i,j+1}={-2\over j+1}a_{i-1,j}=\cdots =
{(-2)^{j+1}\over (j+1)!}a_{i-j-1,0}=:{(-2)^{j+1}\over (j+1)!}a_{i-j-1}
\ee
or
\be
<\P_i(z) \P_n(w)>=(z-w)^{-2\D}\sum_{j=0}^i {(-2)^j\over j!}a_{i-j}
(\log(z-w))^j,
\ee
and also we have
\be
<\P_i(z) \P_k(w)>=<\P_{i+k-n}(z) \P_n(w)>\qquad {\rm for }\qquad 
i+k\geq n.
\ee
So for the case of $n$ logarithmic field, we found all the two point 
functions. The interesting points are 

\noindent i)some of the two-point functions become zero.

\noindent ii)some of the two-point functions are logarithmic, and the highest 
power of the logarithm, which occures in the $<\P_n \P_n>$, is $n$.

The most general case is the case where there is more than one Jordanian 
block in the matrix $D$ ,or in other words, there is more than one set of 
logarithmic operators. The dimension of these blocks may be equal or not 
equal. Using the same procedure, one can find that 
\be
<\P_i^I(z)\P_j^J(w)>=\cases{ (z-w)^{-2\D}\sum_{k=0}^{i+j-n}{(-2)^k\over k!}
a^{IJ}_{n-k}[\log (z-w)]^k,&$i+j\geq n$\cr 0,&$i+j<n$\cr}
\ee
where   $I$ and $J$ label the Jordan cells, $n={\rm max }\{ n_I,n_J\}$
and $n_I$ and $n_J$ are the dimensions of the corresponding Jordan cells.
Also note that the conformal dimensions of the cells $I$ and $J$ must be
equal, otherwise the two-point functions are trivially zero.

Now we want to consider the three-point functions of logarithmic fields.
The simplest case is the case where, besides $\P$, only one extra 
logarithmic field $\Si$ exists in the theory. The three-point functions of 
the fields $\P$ are the same as ordinary conformal field theory.
\be
A(z_1,z_2,z_3):=<\P(z_1) \P(z_2)\P(z_3)>={a\over (\xi_1 \xi_2\xi_3)^{\D}}
=:a f(\xi_1,\xi_2,\xi_3),
\ee
where $$\xi_i={1\over 2}\sum_{j,k}\epsilon_{ijk}(z_j-z_k).$$

If one acts the set $\{L_0,L_{\pm 1}\} $ on the three-point function
$<\Si(z_1) \P(z_2)\P(z_3)>:=B(z_1,z_2,z_3)$, the result is an inhomogeneous 
partial differential equation for $B(z_1,z_2,z_3)$ where the inhomogeneous 
part is $A(z_1,z_2,z_3)$. So the form of $B(z_1,z_2,z_3)$ should be as below,
\be
B(z_1,z_2,z_3)=[b+\sum b_i\log \xi_i]f(\xi_1,\xi_2,\xi_3).
\ee
Solving the above mentioned differential equations, we find the parameters 
$b_i$ to be
\be
b_1=-b_2=-b_3=a.
\ee
The final result is 
\be \label{**}
<\Si(z_1) \P(z_2)\P(z_3)>= [b+a\log {\xi_1\over \xi_2\xi_3}]
f(\xi_1,\xi_2,\xi_3)
\ee
If there are two fields $\Si$ in the three-point function, one can write it 
in the following form
\be 
<\Si(z_1) \Si(z_2)\P(z_3)>= [c+\sum c_i\log \xi_i+\sum_{ij}c_{ij} \log\xi_i 
\log\xi_j]f(\xi_1,\xi_2,\xi_3).
\ee
Again the Ward identities can be used to determine the parameters
$c_i$ and $c_{ij}$,
\be 
<\Si(z_1) \Si(z_2)\P(z_3)>= [c-2b\log \xi_3+a 
[(-{\log\xi_1\over \log \xi_2})^2+(\log\xi_3)^2]f(\xi_1,\xi_2,\xi_3).
\ee
Finally, for the correlator of three $\Si$'s we use
\be 
<\Si(z_1) \Si(z_2)\Si(z_3)>= [d+d_1D_1+d_2D_2+d'_2D_1^2+d_3D_3+d'_3D_1D_2
+d'_3D_1^3] f(\xi_1,\xi_2,\xi_3)
\ee
where
\be
D_1:=\log (\xi_1\xi_2\xi_3) 
\ee
\be
D_2:=\log\xi_1\log \xi_2+\log\xi_2\log \xi_3+\log\xi_1\log \xi_3
\ee
\be
D_3=\log \xi_1 \ \log \xi_2 \ \log \xi_3.
\ee
This is the most general symmetric up to third power logarithmic function
of the relative positions.
Using the Ward identities, this three-point function is calculated to be 
\be 
<\Si(z_1) \Si(z_2)\Si(z_3)>= [d-cD_1+4bD_2-bD_1^2+8aD_3-4aD_1D_2+aD_1^3]
f(\xi_1,\xi_2,\xi_3)
\ee

Now there is a simple way to obtain these correlators. 
Remember of the relation between $\P(z)$ and $\Si(z)$
\be
\Si(z)={\d\over \d\D}\P(z).
\ee
The meaning of this relation will be clearer in section 3.
Consider any three-point function which contains the field $\Si$.
This correlator is related to another correlator which has a $\P$ 
instead of $\Si$ according to
\be
<\Si(z_1)A(z_2)B(z_3)> ={\d\over \d\D}<\P(z_1)A(z_2)B(z_3)>,
\ee
To be more exact, the left hand side satisfies the Ward identities if the
correlator of the right hand side does so.
But the three-point function for ordinary fields are known. So it is enough 
to differentiate it with respect to the weight $\D$. Obviously, a 
logarithmic term appears in the result. In this way one can easily obtain 
the above three-point functions. In fact instead of solving certain partial 
differential equations, one can easily differentiate with respect to the 
conformal weight. This method can also be used when there are $n$ logarithmic 
fields. To obtain the three- point function containing the field $\P_i$,
one should write the three-point function, which contains the 
field $\P_0$, and then differentiate it $i$ times with respect to $\D$.  
Note that in the first three-point function, there may be more than 
one field with the same conformal weight $\D$. Then one must treat the 
conformal weights to be independent variables, differentiate with 
respect to one of them, and finally put them equal to their appropriate
value. Second, there are some constants, or unknown functions in the case of 
more than three-point functions, in any correlator. In differentiation 
with respect to a conformal weight, one must treat these {\it formally} as 
functions of the conformal weight as well.
As an example consider 
\be
<\P(z_1) \P(z_2)\P(z_3)>={a\over (\xi_1)^{\D_2+\D_3-\D_1} 
(\xi_2)^{\D_3+\D_1-\D_2}(\xi_3)^{\D_2+\D_1-\D_3}}.
\ee
Differentiate with respect to $\D_1$, and then put $\D_1=\D_2=\D_3$,
and ${\d a\over \d \D_1}=b$. This is (\ref{**}).

This method can be used for any $n$-point function:
\be 
<\P_i(z_1)\cdots A(z_{n-1})B(z_n)> ={\d^i\over {\d\D}^i}<\P_0(z_1)
\cdots A(z_{n-1})B(z_n)>,
\ee
provided one treats the constants and functions of the correlator as 
functions of the conformal weight. Another thing to be noted is that this 
technique does not work for the two point functions. The reason for this is 
that the two point function of two primary fields with different conformal 
weights is zero. So, the two point function is not a well-behaved 
differentiable function of the conformal weights.

\vspace{0.5cm}
\section{OPE Coefficients of General LCFT}

The most general expression for the operator product expansion of ordinary
conformal fields is \cite{bpz}:
\be\label{phd}
\Phi_n(z)\Phi_m(0)=\s_p z^{\de{p}-\de{n}-\de{m}}\;C_{nm}^p\phi_p(z)
\Phi_p(0)
\ee
where
\be\label{phd2}
\phi_p(z)=\s_{\kk} z^{\s k_i}\;\beta_{nm}^{p,\kk}L_{-k_1}\;\cs L_{-k_n}
\ee
Here the coeficients $\b _{nm}^{p\kk}$ are completely determined in terms
of conformal weights and the central charge of the theory. $C^p_{nm}$'s,
however, are not determined just by conformal invariance.
Now, concentrate on a specific value of $p$, and suppose that there are  
two conformal weights $\D_p$, and $\D'_p:=\D_p +\e $, where $\e $ is a 
small number. One can write
\be \ba{ll}
\Phi_n(z)\Phi_m(0)&=\cdots + z^{\de{p}-\de{n}-\de{m}}\hat C_{nm}^p\phi_p(z)
\Phi_p(0)
+ z^{\de{p}-\de{n}-\de{m}+\e }\hat C_{nm}^{p'}\phi_{p'}(z)\Phi_{p'}(0)\cr
&=\cdots + (\hat C_{nm}^p+\hat {C'}_{nm}^p) z^{\de{p}-\de{n}-\de{m}}\phi_p(z)
\Phi_p(0)+\e \hat {C'}_{nm}^p{\d \over \d \D_p} (z^{\de{p}-\de{n}-\de{m}}
\phi_p(z)\Phi_p(0))\cr
\ea
\ee
where
\be \hat C'{}^p_{nm}:=\hat C^{p'}_{nm},\ee
and we have treated $\p_p$ and $\P_p$, formally, as functions of $\D_p$.

Now let $\e$ tend to zero. If the $C$'s are kept finite, the second term 
vanishes and nothing new happens: this is just an ordinary conformal field 
theory. If, on the other hand, one keeps $\e \hat {C'}$ and 
$\hat {C}+\hat {C'}$ finite. It turns out that
\be
\Phi_n(z)\Phi_m(0)=\cdots + \bar C_{nm}^p z^{\de{p}-\de{n}-\de{m}}\phi_p(z)
\Phi_p(0)+C_{nm}^p{\d \over \d \D_p} (z^{\de{p}-\de{n}-\de{m}}
\phi_p(z)\Phi_p(0))
\ee
As $C$ and $\bar C$ are arbitrary, we can define $\bar C$ as the formal 
derivative of $C$ with respect to $\D_p$. Then 
\be\label{X}
\Phi_n(z)\Phi_m(0)=\cdots +{\d \over \d \D_p}[C_{nm}^p 
z^{\de{p}-\de{n}-\de{m}}\phi_p(z)\Phi_p(0)],
\ee
or
\be
\Phi_n(z)\Phi_m(0)=\cdots + z^{\de{p}-\de{n}-\de{m}}\{ C_{nm}^p[\si_p(z)
\Phi_p(0)+\p_p(z) \Si_p(0)+\p_p(z)\P_p(0)\log z]+
{C'}_{nm}^p\phi_p(z)\Phi_p(0)\}
\ee
where we have defined
\be \ba{ll}
\si_p(z)={\dis{\d \p_p(z)\over \d \D_p }}\cr\cr
\Si_p(z)={\dis{\d \P_p(z)\over \d \D_p }}\cr
\ea
\ee
Note that these derivations are formal. There are, of course, conformal 
field theories where the set of conformal weights is discrete, and it may 
seem that there, derivation with respect to the weight is meaningless. What
is done, resembles very much to the case when one knows a function only in 
certain points. One cannot {\it obtain} the derivative of this function. One 
can, however, introduce other (unknown) quantities as the formal derivative 
of this function and use identities (such as Leibnitz's rule) concerning the 
derivation. There remains, of course, the unknown quantities introduced by 
derivation. That is the reason why new quantities ($c$-numbers such as 
$\bar C$ and operators such as $\si_p$ and $\Si_p$) are introduced. The same 
is true also for derivating the correlators: one cannot {\it obtain} the 
derivative of constants appearing in the correlator. They simply introduce 
more constants in the theory. Also note that all we have used is 
conformal invariance, and no specific model has been taken into account.

Now, using the definitions
\be \ba{ll}
&\mid \D_m>:=\P_m(0)\mid 0>\cr\cr
&\mid z,\D_p>:=\p_p(z)\mid \D_p>\cr\cr
&\qquad =:\s_N z^N\mid N,\D_p>\cr\cr
&\mid {\D '}_p>:={\d \over \d \D_p} \mid \D_p>\cr\cr
&\mid z,{\D '}_p>:={\d \over \D_p}\mid z, \D_p>\cr
\ea \ee
it is seen that 
\be
\mid z,{\D '}_p>=\p_p(z)\mid {\D '_p}>+\si_p(z)\mid \D_p>,
\ee
and
\be
\P_n(z)\mid \D_m>=\s_N z^{\D_p-\D_m-\D_n+N}[ C_{nm}^p(\mid N,{\D '}_p>
+\log z\mid N,\D_p>)+{C'}_{nm}^p\mid N,\D_p>]
\ee
Acting on both side of this relation by $L_j$, one gets
$$\s_N z^{\D_p-\D_m-\D_n+N}[ C_{nm}^pL_j(\mid N,{\D '}_p>
+\log zL_j\mid N,\D_p>)+{C'}_{nm}^pL_j \mid N,\D_p>]
=\s_N z^{\D_p-\D_m-\D_n+N+j}\times$$
\be
\times\{ C_{nm}^p[(\D_p-\D_m+j\D_n+N)(\mid N,{\D '}_p>
+\log z\mid N,\D_p>)+\mid N, \D_p>] 
+ {C'}_{nm}^p(\D_p-\D_m+j\D_n+N)\mid N,\D_p>\}
\ee
using the independency of $z^k$ and $z^k\log z$, it is seen that 
\be
\cases{L_j\mid N+j,\D_p>=(\D_p-\D_m+j\D_n+N)\mid N,\D_p>\cr
L_j\mid N+j,{\D '}_p>=(\D_p-\D_m+j\D_n+N)\mid N,{\D '}_p>
+\mid N,\D_p>\cr},\quad j>0
\ee
The last relation is obviously the derivative of the first, with respect to
$\D_p$.

Similarly, the action of $L_0$ yields
\be \ba{l}
L_0(\mid N,\D_p>=(\D_p+N)\mid N,\D_p>\cr
L_0\mid N,{\D '}_p>=(\D_p+N)\mid N,{\D '}_p>+\mid N,\D-p>.\cr
\ea
\ee
In \cite{rr1}, a method was proposed to obtain $\mid N,\D_p>$ in terms of 
$\mid \D_p>$:
\be
\mid N,\D_p>=\s_{\kk_{\mid \s k_i=N}}\b_{nm}^{p \kk}L_{-k_1}\;\cs L_{-k_n}
\mid \D_p>
\ee
where $\b$'s satisfy a linear equation of the form
\be \label{I}
M\b =\g
\ee
Using the definition of $\mid N,{\D '}_p>$, it is obvious that
\be \label{II}
\mid N,\D'_p>=\s_{\kk_{\mid \s k_i=N}}{\b '}_{nm}^{p,\kk}L_{-k_1}
\;\cs L_{-k_n}\mid \D_p>+\s_{\kk_{\mid \s k_i=N}}\b_{nm}^{p \kk}L_{-k_1}
\;\cs L_{-k_n}\mid \D_p>
\ee
where ${\b '}$'s are the derivative of $\b$'s with respect to $\D_p$.
But taking the derivative of (\ref{I}) yields
\be
M'\b +M {\b '}={\g '}
\ee
Combining this with (\ref{I}), we get 
\be\label{XX}
\pmatrix{ M & 0 \cr M' & M \cr }\pmatrix{\b \cr {\b '}\cr}=
\pmatrix{\g \cr {\g '}\cr}
\ee
which, among with (\ref{II}), it is precisely what was obtained in 
\cite{22}.

Now, the general form of (\ref{X}) can be written as 
\be \P_n(z)\P_m(0)=\s_p\left({\d\over\d\D p}\right)^{q_p}\left[ C^p_{nm}
z^{\D_p-\D_n-\D_m}\p_p(z)\P_p(0)\right],\ee
where $q_p$ is the dimension of the $p$th. Jordanian block.

In a manner similar to that of the previous discussion, one can define
\be\cases{\P_p^{(n)}:=\left({\d\over\d\D p}\right)^n\P_p\cr
\p_p ^{(n)}:=\left({\d\over\d\D p}\right)^n\p_p\cr
\mid N,D_p^{(n)}>:=\left({\d\over\d\D p}\right)^n\mid N,\D_p>}
\qquad ,0\leq n<q_p\ee
and the equation corresponding to (\ref{XX}) becomes
\be\s^{q_p-1}_{j=0}M_{ij}\b_j=\g_i,\ee
where
\be\g_i:=\left({\d\over\d\D p}\right)^i\g,\ee
and
\be M_{ij}:={{i!}\over{j!(i-j)!}}\left({\d\over\d\D p}\right)^{i-j}M.\ee
M's up to the third level are given in \cite{rr1}.
\\ 

{\bf Acknowledgement} M. K. would like to thank the research 
vice-chancellor of Tehran University, this work was partially supported by 
them.

 \newpage

 %%%%%%%%%%%%%%%%%%%%%%%%%%%%%%%%%%


\begin{thebibliography}{99}
\bibitem{gu} V. Gurarie, Nucl. Phys. {\bf B410}[FS] (1993) 535.
\bibitem{srz} L. Rozansky and H. Saleur, Nucl. Phys. {\bf B376} (1992) 461.
\bibitem{5}  X.G. Wen, Y.S. Wu and Y. Hatsugai, Nucl. Phys. {\bf B422}[FS]
(1994) 476.
\bibitem{rr1} M.R. Rahimi Tabar and S. Rouhani, Annals of Phys. {\bf 246} 
(1996)446.
\bibitem{rr2} M.R. Rahimi Tabar and S. Rouhani, ``The Alf'ven Effect and
Conformal Field Theory'' IPM Preprint IPM-094-95 [hep-th/9507166] to appear 
in Nuovo Cimento {\bf B}
\bibitem{rr3} M.R. Rahimi Tabar and  S. Rouhani, Europhys. Lett {\bf 37} 447 (1997)
\bibitem{flo} M. A. I. Flohr ``2-Dimensional Turbulence: yet another
Conformal Field Theory Solution'' IASSNS-HEP-96/69,hep-th 9606130
\bibitem{rahim1} M.R. Rahimi Tabar and  S. Rouhani  Phys. Lett. {\bf A224} 331 (1997)
\bibitem{cp1} M. R. Gaberdiel, H.G. Kausch, Nucl.Phys. {\bf B447} (1996)
293.
\bibitem{bk} A. Bilal and I.I. Kogan, Princton Preprint PUPT-1482
[hep-th/9407151]; Nucl. Phys. {\bf B449} (1995) 569.
\bibitem{b16} I.I. Kogan and N.E. Mavromatos, Phys. Lett. {\bf B375} (1996)
111.
\bibitem{10} J.S. Caux, I.I. Kogan and A.M. Tsvelik,
Nucl.Phys.{\bf B466} (1996) 444.
\bibitem{d14}Z. Maassarani, D. Serban,`` Non-Unitary Conformal Field Theory
and Logarithmic Operators for Disordered Systems'', preprint SPHT-T96/037
(1996), hep-th/9605062
\bibitem{W} R. Blumenhagen, W. Eholzer, A. Honecker, K. Hornfeck, R. Hubel,
Phys. Lett. {\bf B332} (1994)51, Int. J. Mod. Phy. {\bf A10} (1995) 2367.
\bibitem{b17} J. Ellis, N.E. Mavromatos and D.V. Nanopoulos,`` D Branes from
Liouville Strings'' CERN-TH/96-81, hep-th/9605046
\bibitem{fl} M.A.I. Flohr, Int. J. Mod. Phys. {\bf A11} (1996) 4147.
\bibitem{fufl} M.A.I. Flohr, ``On Fusion Rules in Logarithmic Conformal
Field Theories'' IASSNS-HEP-96/54, hep-th/9605151
\bibitem{bpz}A.M. Belavin, A.M. Polyakov and A.B. Zamolodchikov, Nucl. Phys.
 {\bf B241} (1984) 333.
\bibitem{22} A. Shafiekhani, M.R. Rahimi Tabar, hep-th/9604007, to appear in 
Int. J. Mod. Phys. {\bf A}.
\end{thebibliography}
\end{document}